\documentclass[useAMS,usenatbib]{mn2e}
\usepackage{savesym}
\usepackage{txfonts}
\savesymbol{iint}
\savesymbol{iiint}
\savesymbol{iiiint}
\savesymbol{idotsint}
\usepackage{graphicx}
\usepackage{amssymb}
\usepackage[below]{placeins}
\usepackage{txfonts}
\usepackage{natbib}
\usepackage{amsmath}
\usepackage{float}
\restoresymbol{TXF}{iint}
\restoresymbol{TXF}{iiint}
\restoresymbol{TXF}{iiiint}
\restoresymbol{TXF}{idotsint}
\usepackage{multirow}
\usepackage{array}
\usepackage{xcolor}
\usepackage[percent]{overpic}
\bibpunct{(}{)}{;}{a}{}{,}

\def\spose#1{\hbox to 0pt{#1\hss}}
\def\approxlt{\mathrel{\spose{\lower 3pt\hbox{$\sim$}}
        \raise 2.0pt\hbox{$<$}}}
\def\approxgt{\mathrel{\spose{\lower 3pt\hbox{$\sim$}}
        \raise 2.0pt\hbox{$>$}}}
\def\approxpropto{\mathrel{\spose{\lower 3pt\hbox{$\sim$}}
        \raise 2.0pt\hbox{$\propto$}}}
\mathchardef\twiddle="2218

\def\multleft#1{\hbox to size{\vbox {\halign {\lft{##}\cr #1}}\hfill}\par}
\def\multright#1{\hbox to size{\vbox {\halign {\rt{##}\cr #1}}\hfill}\par}

\def\today{\ifcase\month\or January\or February\or March\or April\or May\or
      June\or July\or August\or September\or October\or November\or December\fi
      \space\number\day, \number\year}
\def\<{\thinspace}

\def\ga{{\rm\thinspace gauss}}

\makeatletter
\newcommand{\thickhline}{%
    \noalign {\ifnum 0=`}\fi \hrule height 1.2pt
    \futurelet \reserved@a \@xhline
}
\newcolumntype{"}{@{\hskip\tabcolsep\vrule width 1pt\hskip\tabcolsep}}
\makeatother

\newcommand{\ion}[2]{#1\,{\sc{#2}}}

\bibpunct{(}{)}{;}{a}{}{,}

\title[Studying the hot coronae of massive relic galaxies]{Digging for red nuggets: discovery of hot halos surrounding massive, compact, relic galaxies}

\author[Werner et al.]{N. Werner$^{1,2,3}$\thanks{wernernorbi@gmail.com}, K. Lakhchaura$^{1}$, R. E. A. Canning$^{4}$\thanks{{\it Einstein} Fellow}, M. Gaspari$^{5}$\thanks{{\it Einstein} and {\it Spitzer} Fellow}, A. Simionescu$^6$  \\
$^1$MTA-E\"otv\"os University Lend\"ulet Hot Universe and Astrophysics  Research Group, P\'azm\'any P\'eter s\'et\'any 1/A, Budapest, 1117, Hungary \\
$^2$Department of Theoretical Physics and Astrophysics, Faculty of Science, Masaryk University, Kotl\'a\v{r}sk\'a 2, Brno, 611 37, Czech Republic \\
$^3$School of Science, Hiroshima University, 1-3-1 Kagamiyama, Higashi-Hiroshima 739-8526, Japan \\
$^4$Kavli Institute for Particle Astrophysics and Cosmology, Stanford University, 452 Lomita Mall, Stanford, CA 94305-4085, USA \\
$^5$Department of Astrophysical Sciences, Princeton University, 4 Ivy Lane, Princeton, NJ 08544-1001, USA \\
$^6$Institute of Space and Astronautical Science (ISAS), JAXA, 3-1-1 Yoshinodai, Chuo-ku, Sagamihara, Kanagawa, 252-5210, Japan\\
}
\begin{document}
\maketitle

\begin{abstract}
We present the results of {\it Chandra} X-ray observations of the isolated, massive, compact, relic galaxies MRK~1216 and PGC~032873. Compact massive galaxies observed at $z>2$, also called red nuggets, formed in quick dissipative events and later grew by dry mergers into the local giant ellipticals. Due to the stochastic nature of mergers, a few of the primordial massive galaxies avoided the mergers and remained untouched over cosmic time. We find that the hot atmosphere surrounding MRK~1216 extends far beyond the stellar population and has an 0.5--7~keV X-ray luminosity of $L_{\rm X}=(7.0\pm0.2)\times10^{41}$~erg~s$^{-1}$, which is similar to the nearby X-ray bright giant ellipticals. The hot gas has a short central cooling time of  $\sim50$~Myr and the galaxy has a $\sim13$~Gyr old stellar population. The presence of an X-ray atmosphere with a short nominal cooling time and the lack of young stars indicate the presence of a sustained heating source, which prevented star formation since the dissipative origin of the galaxy 13 Gyrs ago. The central temperature peak and the presence of radio emission in the core of the galaxy indicate that the heating source is radio-mechanical AGN feedback. Given that both MRK~1216 and PGC~032873 appear to have evolved in isolation, the order of magnitude difference in their current X-ray luminosity could be traced back to a difference in the ferocity of the AGN outbursts in these systems.
Finally, we discuss the potential connection between the presence of hot halos around such massive galaxies and the growth of super/over-massive black holes via chaotic cold accretion.
\end{abstract}


\section{Introduction}

The formation and evolution of giant elliptical galaxies is well described by two-phase models \citep[][]{oser2010,rodriguez2016}. The first phase is a quick dissipative event, when the core of the galaxy and its supermassive black hole are formed. The results of this first stage are the compact massive galaxies with $r_{\rm e} \lesssim 2$~kpc and $M_{\star}\gtrsim10^{11}~M_{\odot}$, so called red nuggets observed at $z>2$. This early rapid growth is followed by a slow accretion phase when the galaxy undergoes dry mergers with lower mass galaxies. These random encounters will place most of the newly accreted material at the periphery of the galaxy, significantly increasing its size, but leaving the centre unaffected. Semi-analytical models and cosmological simulations indicate that the size of a massive galaxy can increase by a factor of $\sim7$  during the merger phase, while its velocity dispersion increases by at most a factor $\sim1.1$ \citep{hilz2012}. 
However, due to the stochastic nature of mergers, a few of the primordial massive galaxies should avoid the second stage, remaining untouched over cosmic time \citep{quilis2013}.

\begin{figure*}
\begin{center}
\begin{minipage}{0.32\textwidth}
\includegraphics[width=1.05\textwidth,clip=t,angle=0.]{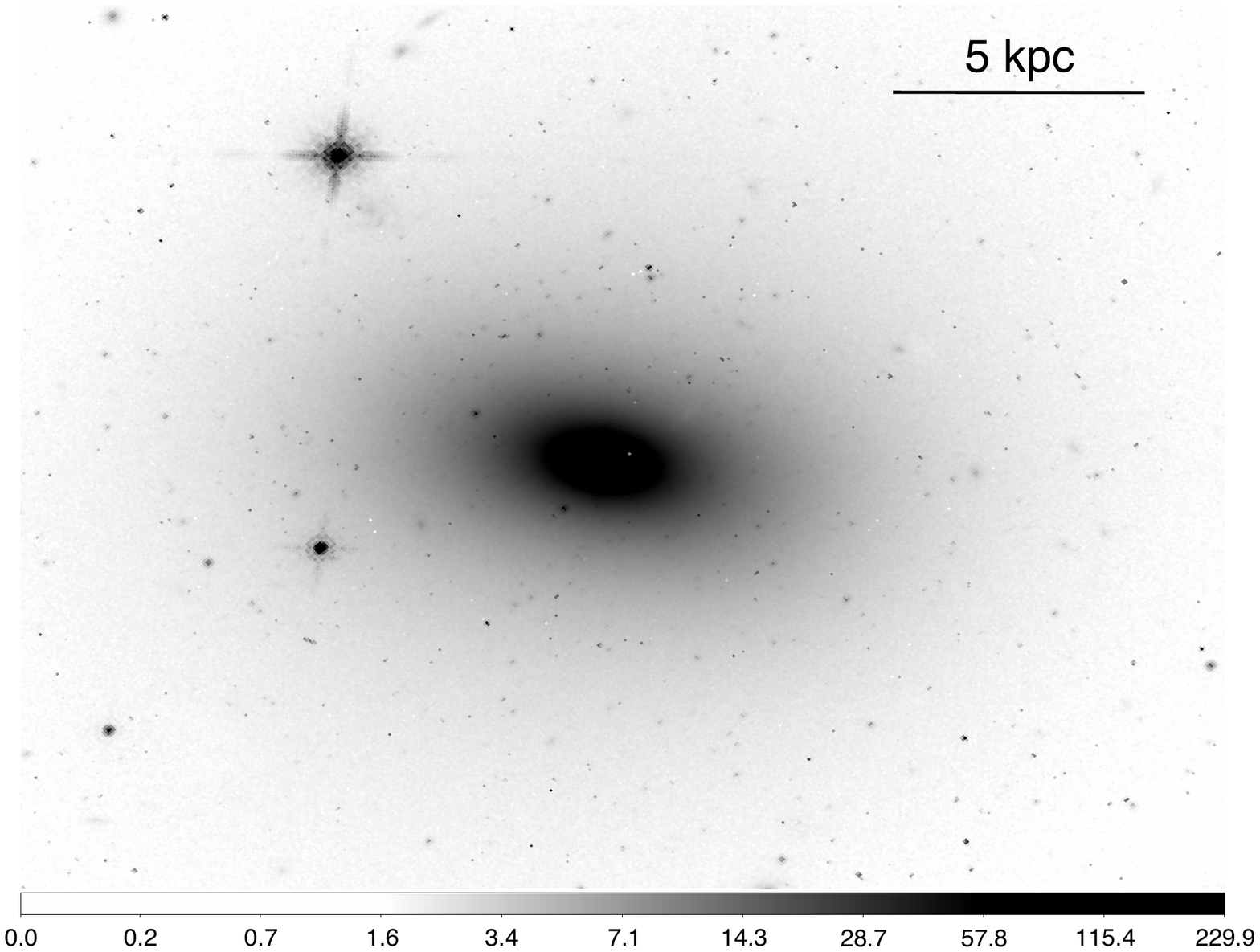}
\end{minipage}
\begin{minipage}{0.32\textwidth}
\hspace{0.2cm}\includegraphics[width=1.05\textwidth,clip=t,angle=0.]{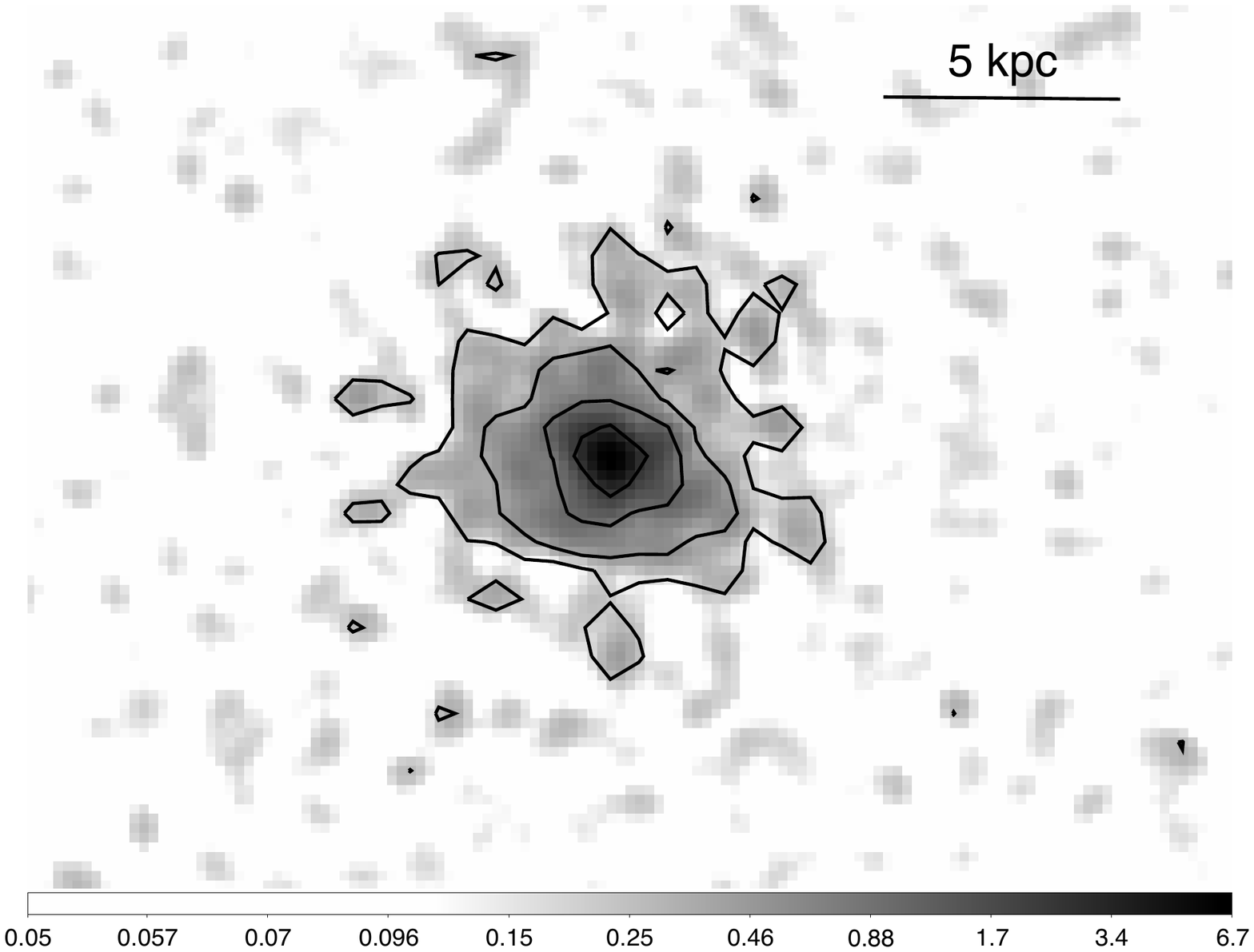}
\end{minipage}
\begin{minipage}{0.32\textwidth}
\vspace{-0.5cm}
\hspace{0.4cm}\includegraphics[width=1.08\textwidth,clip=t,angle=0.]{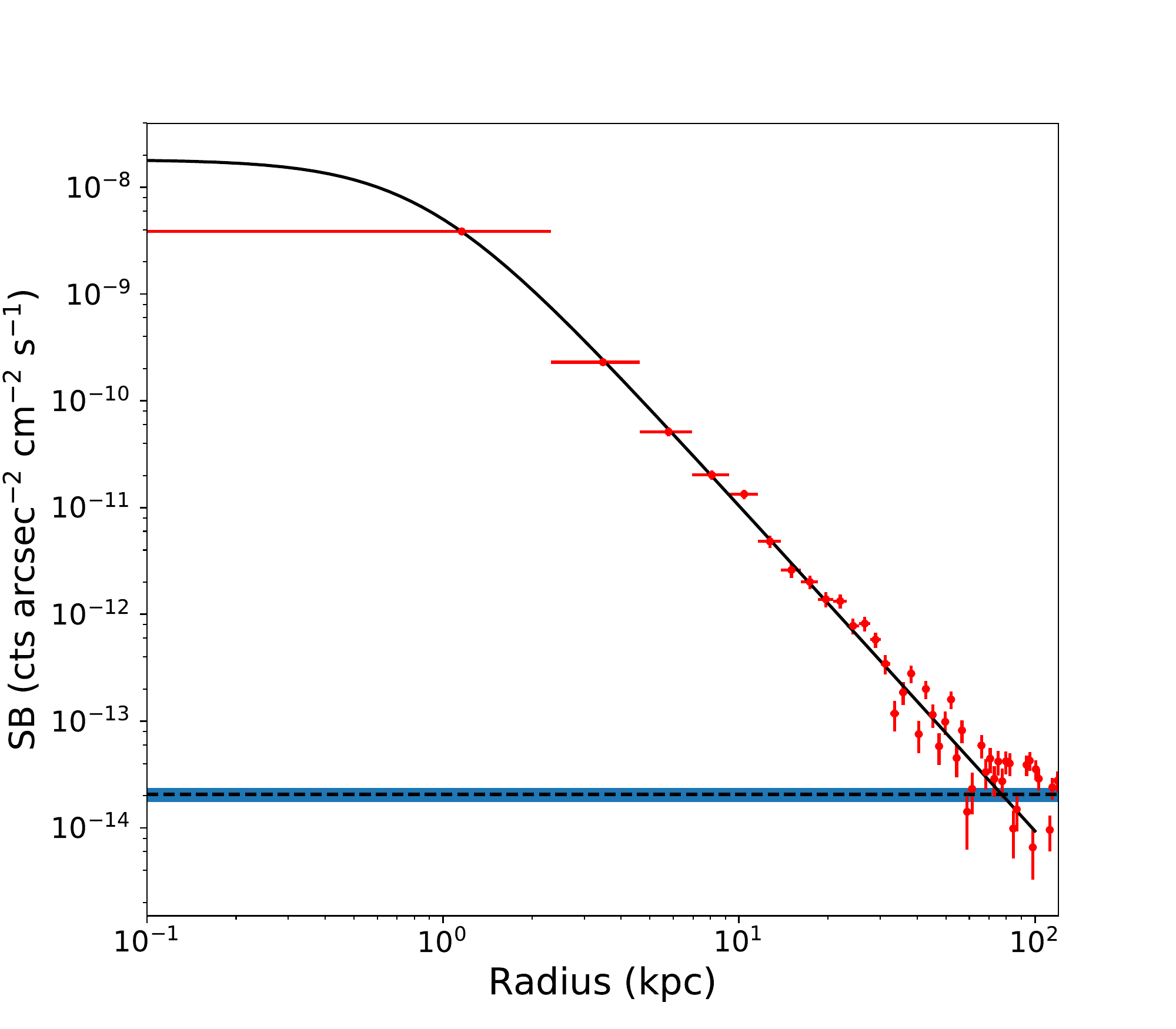}
\end{minipage}
\end{center}
\caption{{\it Left panel:} Hubble Space Telescope (HST) Advanced Camera for Surveys filter F160W archive image of MRK~1216. For a circularised stellar mass density profile see \citet{ferre2017}. {\it Central panel:} The {\it Chandra} X-ray data reveal a hot X-ray emitting atmosphere around the isolated, massive, compact, relic galaxy Mrk 1216. {\it Right panel:} The X-ray surface brightness profile in the 0.5--7.0~keV band shows that the X-ray emitting atmosphere of this relic galaxy extends far beyond its stellar population, out to radii of over $55$~kpc. The emission is fit with a beta model of $\beta=0.66\pm0.15$ and $r_{\rm c}=0.7\pm0.1$~kpc.}
\label{images}
\end{figure*}

The first confirmed low redshift {\it massive relic galaxy}, mimicking the properties of high-redshift compact massive galaxies, is NGC~1277 ($r_{\rm e}=1.2$ kpc and $M_{\star}=1.3\times10^{11}~M_{\odot}$) in the Perseus cluster \citep{trujillo2014}. This galaxy is also well known for hosting one of the most massive black holes detected to date \citep{vandenBosch2012,graham2016,walsh2016}. \citet{ferre2015} identified a sample of seven potential massive relic galaxies, all with unusually massive central black holes (3--$5\sigma$ outliers on the $M_{\rm BH}-M_{\rm bulge}$ relation). Recently, \citet{ferre2017} confirmed that two previously identified candidates (MRK~1216 and PGC~032873) are indeed ``red nuggets" in the present day Universe \citep[see also][]{walsh2017}. The stellar masses of these galaxies reach $\sim2\times10^{11}~M_{\odot}$ and their stellar populations are highly concentrated in the innermost parts, resulting in effective radii of $R_{\rm e}\sim 2$~kpc. \citet{ferre2017} conclude that these galaxies were formed quickly and early, and their mean mass-weighted ages are $\sim13$ Gyr. They have strongly peaked velocity dispersion profiles with $\sigma\sim360$~km~s$^{-1}$ at $R_{\rm e}$ and compact morphologies with no signs of interactions. These properties set them clearly apart from typical giant ellipticals; instead, they represent the properties of the early population of red nuggets observed at redshifts $z\gtrsim2$ \citep[e.g.][]{buitrago2008, vanderwel2011}, that only went though the first dissipative phase of galaxy formation. These systems thus allow us to perform a detailed study of red nuggets, the puzzling progenitors of giant elliptical galaxies.

As progenitors of giant elliptical galaxies, red nuggets are expected to have large total masses of $10^{12-13}~M_{\odot}$ and thus, in principle, to hold on to hot X-ray emitting atmospheres. 
The presence of hot atmospheres around massive galaxies in the early universe would have important consequences for studies of galaxy quenching and maintenance mode feedback. 
The X-ray morphologies, thermodynamic properties, and metallicities of these atmospheres will also carry important information about the more recent growth and evolution of these systems. 
The most isolated of the currently confirmed low redshift massive, compact, relic galaxies are MRK~1216 ($z=0.021328$, $D=97$~Mpc) and PGC 032873 ($z=0.024921$, $D=108$~Mpc), with their closest neighbours at distance $\gtrsim1$~Mpc \citep{ferre2017,yildirim2017}. However, while PGC 032873 is X-ray faint and its 22.7~ks archival {\it Chandra} observation only provides 200 counts, which only allow for a temperature measurement \citep[see Section \ref{analysis} and][]{buote2017}, MRK~1216 is X-ray bright and allows to perform a detailed study of the X-ray emitting atmosphere of the galaxy. Therefore, this massive relic system provides the best, albeit indirect, opportunity to study the thermodynamic structure of an extended X-ray emitting halo around a red nugget, long before future large X-ray missions, such as {\it Athena} or {\it Lynx}, will allow us to observe the massive high redshift galaxies directly.

 \section{Observations and data analysis}
\label{analysis}

MRK~1216 has been observed by {\it Chandra} for 12.9~ks in June 2015 and PGC 032873 for 22.7~ks in March 2015 using the Advanced CCD Imaging Spectrometer (ACIS) chip 6. We analysed these archival data using standard data analysis procedures \citep[e.g.][]{lakhchaura2016,werner2014,werner2016b}. The background, both for image analysis and  spectroscopy, was extracted from the same chip as the source spectrum.

\begin{figure*}
\begin{center}
\begin{minipage}{0.49\textwidth}
\includegraphics[width=1\textwidth,clip=t,angle=0.]{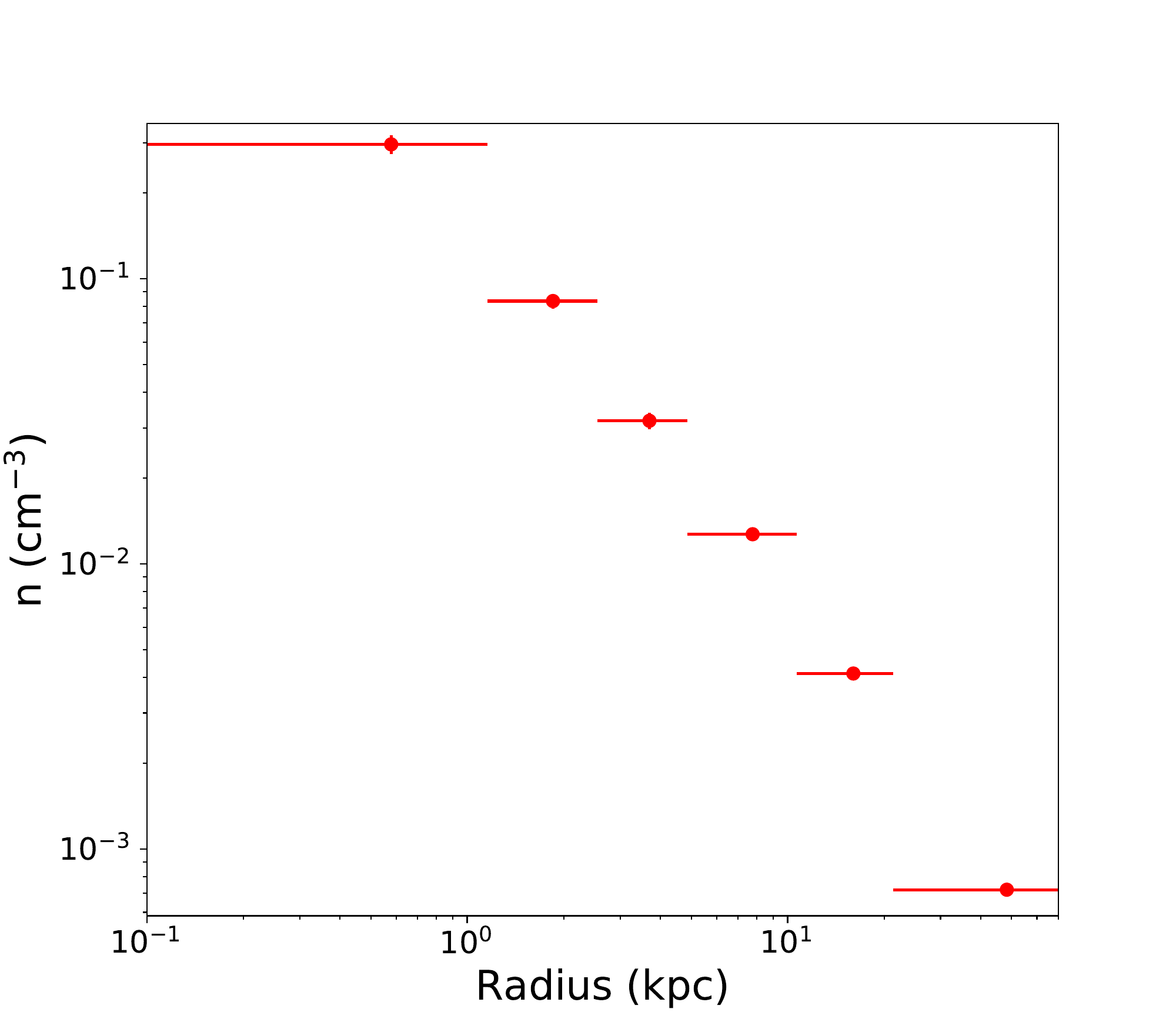}
\includegraphics[width=1\textwidth,clip=t,angle=0.]{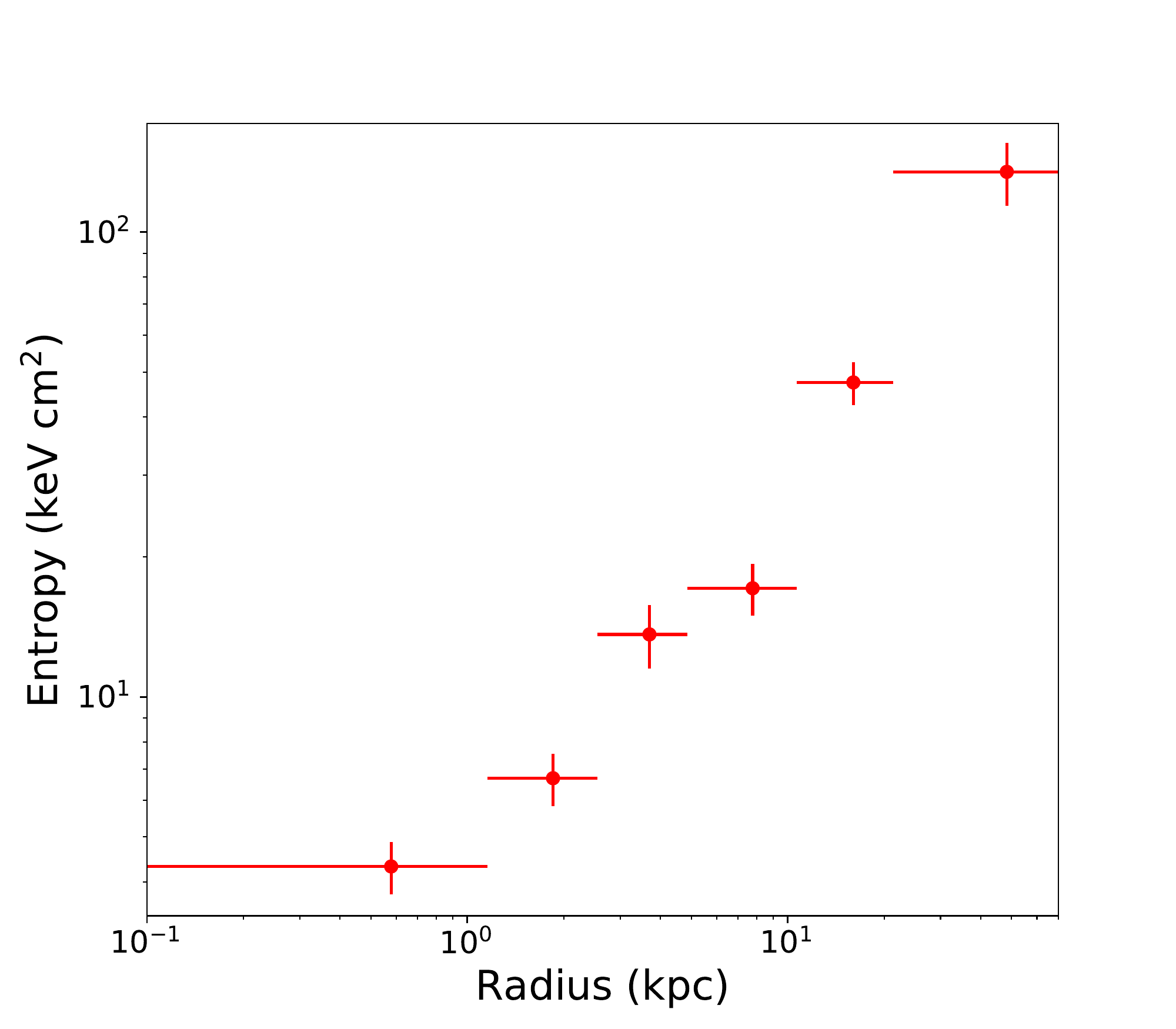}
\end{minipage}
\begin{minipage}{0.49\textwidth}
\includegraphics[width=1\textwidth,clip=t,angle=0.]{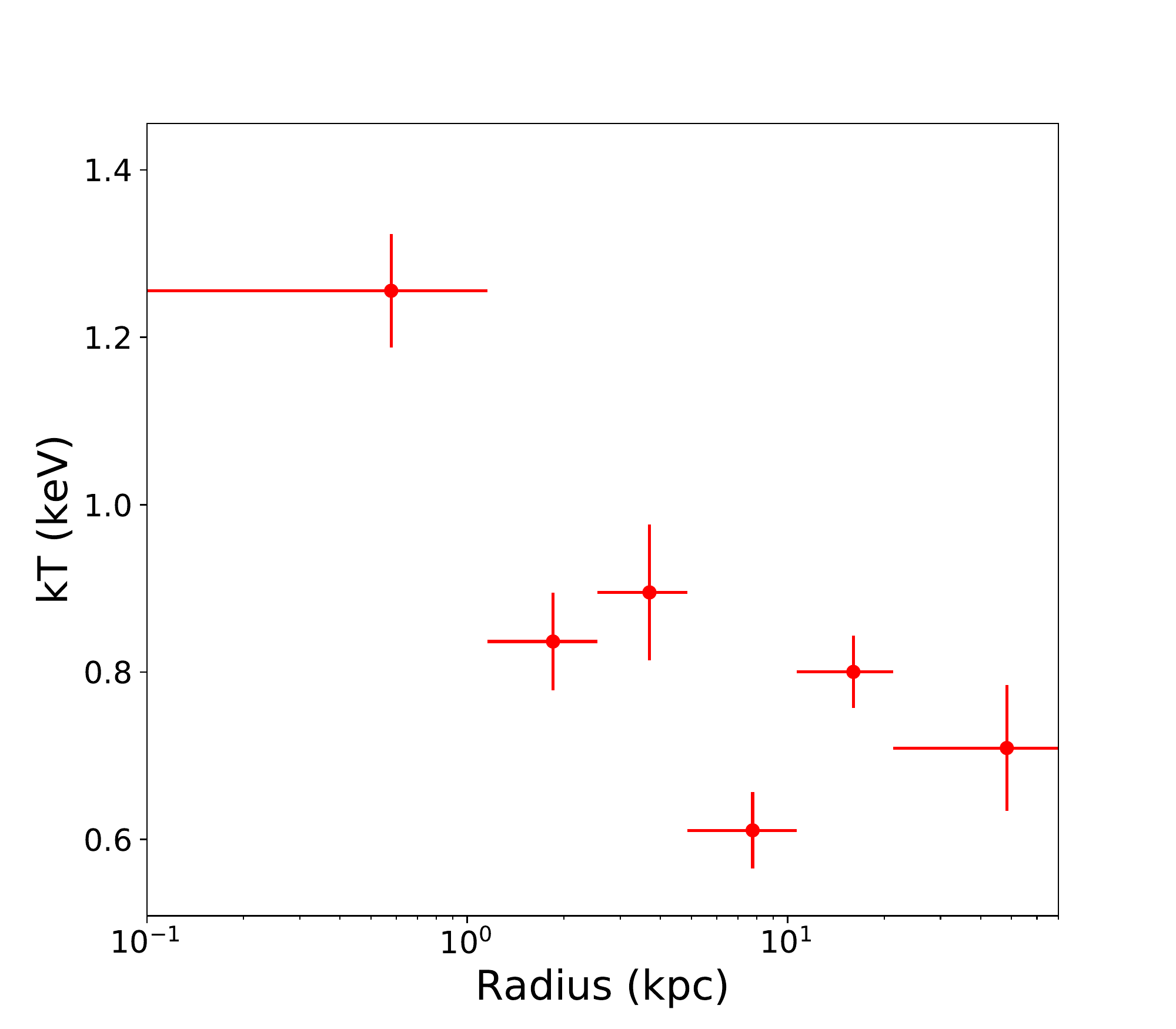}
\includegraphics[width=1\textwidth,clip=t,angle=0.]{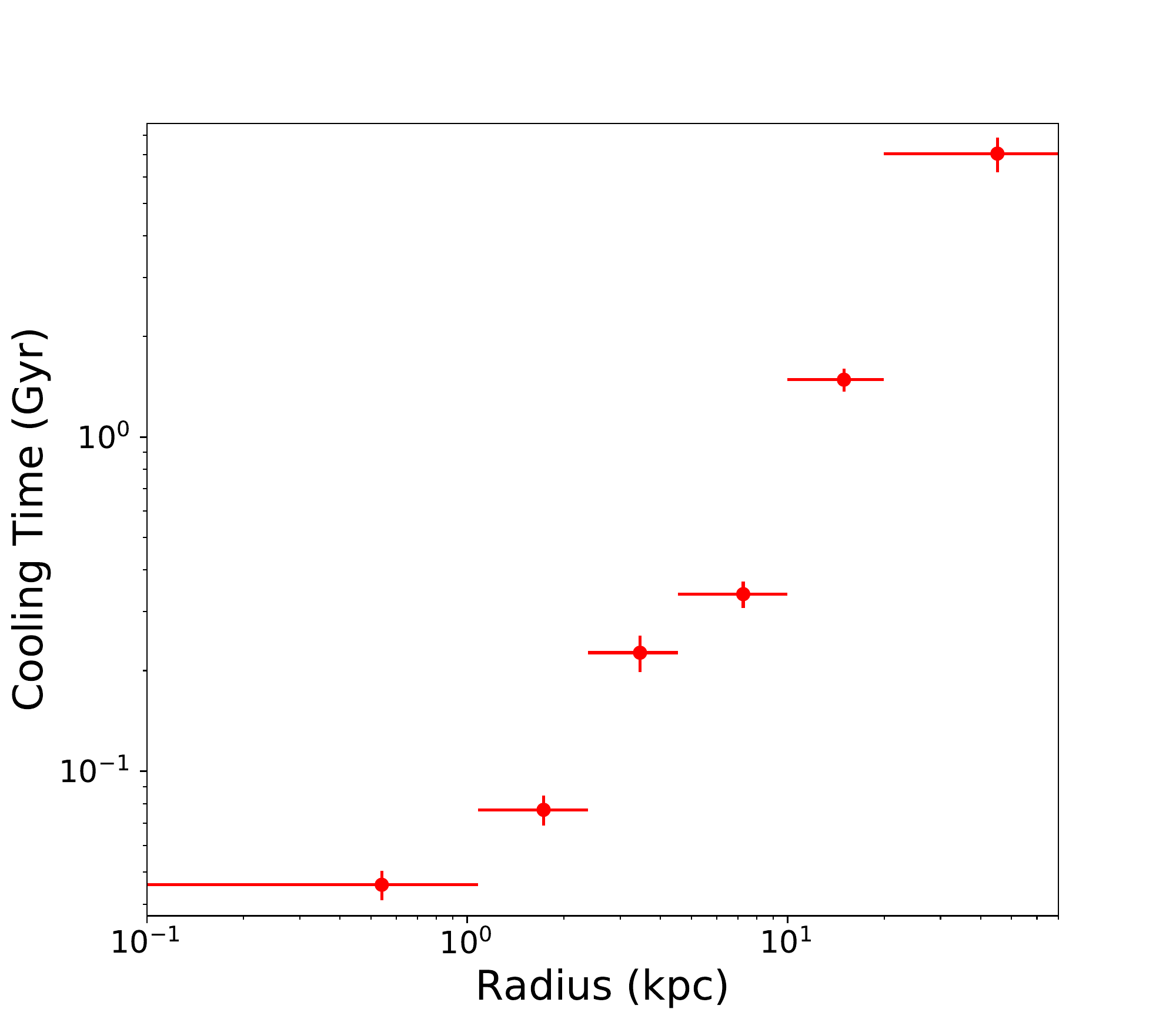}
\end{minipage}
\end{center}
\caption{Radial distributions of the deprojected thermodynamic properties of the X-ray emitting atmosphere of the massive relic galaxy Mrk~1216: total density $n=n_{\mathrm{e}}+n_{\mathrm{i}}$  (top left), temperature  (top right), entropy (bottom left), and cooling time (bottom right).  }
\label{profiles}
\end{figure*}

We modelled the spectra as absorbed single-phase plasma in collisional ionisation equilibrium \citep[APEC, AtomDB 3.0.9][]{smith2001,foster2012}, with the temperature, spectral normalisation and metallicity as free parameters, and a $kT=7.3$~keV bremsstrahlung model with a free normalisation to account for the population of unresolved galactic point sources \citep{irwin2003,boroson2011}. 
The normalisation of the unresolved point source component has large errors and is not statistically significant for these galaxies \citep[see also][]{buote2017}. The line-of-sight absorption column density was fixed to $N_{\rm H}=4.03\times10^{20}$~cm$^{-2}$ for MRK~1216 and $N_{\rm H}=1.1\times10^{20}$~cm$^{-2}$ for PGC 032873, as determined by the Leiden/Argentine/Bonn radio survey of \ion{H}{i} \citep{kalberla2005}. For fitting multi-temperature spectra we used the SPEX spectral fitting package \citep{kaastra1996}.

The spectrum of PGC~032873 contains around 200 counts and  only allows to determine the X-ray luminosity of the gaseous atmosphere of the galaxy and a single temperature value (see Sect. \ref{results}). It is not sufficient to study the spectral properties in multiple radial bins. The data for the significantly brighter MRK~1216 allow to determine the azimuthally averaged deprojected thermodynamic quantities for the X-ray emitting halo surrounding the system in six annular regions, with $\sim200$ background-subtracted counts per annulus (the outermost annulus contains $\sim 300$ counts).
We modelled the spectra using the PROJCT model implemented in the XSPEC spectral fitting package \citep{arnaud1996}. The combined set of spectra was modelled in the 0.6--7.0 keV band simultaneously to determine the deprojected electron density ($n_{\rm e}$) and temperature ($kT$) profiles. From the electron densities and temperatures we determined the entropy, $K=kT_{\rm e}/n_{\rm e}^{\frac{2}{3}}$, and 
cooling time, $t_{\mathrm{cool}}=\frac{3}{2}(n_{\mathrm{e}}+n_{\mathrm{i}})kT/(n_{\mathrm{e}}n_{\mathrm{i}}\Lambda(T)$), profiles, where the ion number density $n_{\rm i} = 0.92n_{\rm e}$, and  $\Lambda(T)$ is the cooling function for Solar metallicity tabulated by \citet{schure2009}.

\begin{table*}
\caption{Deprojected thermodynamic properties of MRK~1216.}
\label{table1}
\centering
\begin{tabular}{cccccccc}
\hline\hline
Radial range  		&  $n$ 				& $kT$ 			& $K$ 			&  $P$				 & $t_{\rm cool}$ 	& $t_{\rm ff} $		& $t_{\rm cool}/t_{\rm ff} $  	\\
	(kpc)			& (cm$^{-3}$)			&  (keV)			& (keV cm$^2$)	& ($10^{-11}$ erg cm$^{-3}$)& (Gyr)			& (Myr)			&  					 \\
\hline 
0--1.2			&  $0.296\pm0.022$ 	 	& $1.25\pm0.07$	& $4.3\pm0.6$		& $59\pm6$  			& $0.052\pm0.005$  & $1.52\pm0.07$ 	&  $34\pm4$	\\
1.2--2.5			&  $0.084\pm0.005$ 	 	& $0.84\pm0.06$  	& $6.7\pm0.9$		& $11.2 \pm 1.0$		& $0.086\pm0.009$	& $6.0\pm0.3$  	&  $15\pm2$	\\
2.5--4.9 			&  $0.032\pm 0.002$ 	& $0.90\pm0.08$  	& $13.6\pm2.1$	& $4.5\pm 0.5$			& $0.26\pm0.03$ 	& $12.1\pm0.6$	&  $21\pm3$	\\
4.9--10.7 			&$0.0143\pm 0.0007$	&$0.61\pm0.05$  	& $17.1\pm2.2$	& $1.24\pm0.11$  		& $0.38\pm0.04$	& $27.4\pm1.3$	&  $14\pm1$	\\
10.7--21.3			&$0.0041\pm 0.0002$	&$0.80\pm0.04$  	& $48\pm5$		& $0.53\pm0.04$		& $1.67\pm0.13$	& $54\pm3$		&  $31\pm3$	\\
21.3--75.2 		&$0.0007\pm0.0001$ 	&$0.71\pm0.08$  	& $135\pm21$		& $0.082\pm0.010$		& $7.9\pm0.9$		& $151.\pm10$		&  $53\pm7$	\\
\hline
\label{obs}
\end{tabular}\\
\end{table*}

\section{Results}
\label{results}

The {\it Chandra} X-ray observation revealed a gaseous X-ray halo surrounding MRK~1216, extending far beyond its stellar population (see Fig.~\ref{images}). The X-ray emission is detected with a significance greater than 3$\sigma$ out to a radius of 55~kpc. The spectrum of the central region is entirely consistent with thermal emission. The upper limit on the 0.5--7~keV X-ray luminosity of a power-law like emission component from the active galactic nucleus (AGN) and possibly unresolved point sources within the radius of 1 kpc is $9.47\times10^{39}$~erg~s$^{-1}$, which is about 3.4 per cent of the thermal emission in this region. The total 0.5--7~keV X-ray luminosity within the radius of 10~kpc is $L_{\rm X}=(7.0\pm0.2)\times10^{41}$~erg~s$^{-1}$. Around PGC~032873, {\it Chandra} X-ray  observations revealed a hot halo extending only to $r\sim20$~kpc \citep[10 arcsec; see also][]{buote2017} with a total 0.5--7~keV X-ray luminosity within $r=10$~kpc of $L_{\rm X}=(5.6\pm0.5)\times10^{40}$~erg~s$^{-1}$.

By fitting a model of a single-temperature plasma in collisional ionisation equilibrium to the spectrum of MRK~1216 extracted within the radius of $r=10$ kpc, we determine a best fit temperature of $kT=0.80\pm0.03$~keV and a metallicity of $Z=0.6\pm0.1$~Solar \citep[consistent values are obtained using both XSPEC/APEC and the SPEX spectral fitting packages; we have assumed the Solar abundances of ][thoughout]{lodders2009}. This metallicity is likely to be biased low due to an intrinsically multi-temperature structure of the gas \citep[see][]{buote2000, werner2008}. When fitting the spectrum with a Gaussian emission measure model (GDEM) available in SPEX  \citep[see e.g.][for details]{deplaa2017}, we find a multi-temperature distribution with a central temperature of $0.92\pm0.9$~keV, width $\sigma=0.3\pm0.1$~keV and a metallicity of $1.2\pm0.3$~Solar. For the spectrum of PGC~032873 extracted within $r=10$~kpc we find a temperature of $kT=0.55\pm0.11$~keV. The low number of counts does not allow us to measure the metallicity of this system. 

The data for MRK~1216 allow us to determine the radial distribution of azimuthally averaged spectral properties in six concentric annuli. Our best fit deprojected profile shown in Fig.~\ref{profiles} and Table \ref{table1} has a C-statistics value of 1284 for 1842 degrees of freedom.
The galaxy has a relatively high central density of $n=0.30\pm0.02$~cm$^{-3}$, a centrally peaked temperature distribution (core temperature of $kT=1.25\pm0.07$~keV), a relatively flat entropy profile with a power-law index of $0.78\pm0.05$, and a central cooling time of $t_{\rm cool}=52\pm5$ Myr. The short cooling time means that unchecked radiative cooling would lead to reservoirs of cold gas and star formation. If we fit the spectra with a simple isobaric cooling flow model (although in reality the presence of simple isobaric cooling is unlikely), we obtain a mass deposition rate  between 0.1~keV and 1.2~keV of $\dot{M}=1.18\pm0.14~M_{\odot}$~yr$^{-1}$.  Taken together with the old stellar population of the galaxy \citep[mean mass weighted age of $12.8\pm1.5$~Gyr, with 99\% of the stellar population more than 10 Gyrs old;][]{ferre2017}, the short cooling time implies the presence of a heating source, which prevents the radiative cooling of the halo. 
The temperature distribution is strongly centrally peaked  consistent with the presence of a nuclear heating source. The presence of warm/cool ionised gas has not been reported for this system. The Very Large Array (VLA) NVSS survey \citep{condon1998} shows a central radio source in the galaxy, with a flux of $9.2\pm0.6$ mJy at 1.4 GHz, corresponding to a radio luminosity of $1.9\times10^{38}$~erg~s$^{-1}$.  

While the total hot gas mass within 70 kpc is $(1.9\pm0.5)\times10^{10} M_{\odot}$, the gas mass within the smaller radius of $r<10$ kpc is only $M_{\rm gas}=(1.40\pm0.06)\times10^{9}~M_{\odot}$, comparable to the gas mass in nearby giant ellipticals \citep[e.g.][]{werner2014}. The gas mass is a small fraction of the stellar mass, which is $M_{\star}=(2.0\pm0.8)\times10^{11}~M_{\odot}$.  The total mass within the same radius, calculated from the pressure profile assuming hydrostatic equilibrium, is $M_{\rm tot}=(7\pm5)\times10^{11}~M_{\odot}$, consistent with the dynamical mass of the system \citep[see][]{yildirim2015}. The gas mass fraction is also within the range measured in the nearby, mature, massive elliptical galaxies \citep[see e.g.][]{werner2012}. For a more detailed mass modelling of MRK~1216 based on its X-ray properties see \citet{buote2017}. They extrapolated the mass profile out to the virial radius yielding a total mass within $r_{200}$ of $M_{200} = (9.6\pm3.7)\times10^{12}~M_{\odot}$ for this system, which is consistent with the masses of group scale halos.

\section{Discussion}
\label{discussion}

The isolated, massive, compact relic galaxies MRK~1216 and PGC~032873 harbour extended hot X-ray emitting atmospheres. The X-ray gas surrounding  MRK~1216 extends far beyond its stellar population and its luminosity within $r=10$~kpc is $L_{\rm X}=(7.0\pm0.2)\times10^{41}$~erg~s$^{-1}$, which is similar to the luminosities of the hot atmospheres of nearby bright giant ellipticals, most of which are found in the centres of groups of galaxies \citep[e.g.][]{werner2014}. On the other hand, the X-ray luminosity of PGC~032873, which has a similar stellar mass to MRK~1216 \citep{ferre2017}, is about an order of magnitude lower, $L_{\rm X}=(5.6\pm0.5)\times10^{40}$~erg~s$^{-1}$. This luminosity difference is well within the scatter in the X-ray luminosities of early type galaxies \citep[e.g.][]{osullivan2001,su2015,kim2015}. The hydrostatic mass estimates and the stellar velocity dispersions do not indicate that the gravitational potential well of PGC 032873 is significantly shallower than that of Mrk~1216.

Hot atmospheres might be leftover material from the process of galaxy formation and stellar mass loss may also have contributed significantly to the X-ray emitting gas mass \citep[for a review see][]{mathews2003}. Hydrodynamic simulations predict that about 75 per cent of the ejecta produced by red giant stars moving supersonically relative to the ambient medium will be shock heated to approximately the temperature of the hot gas \citep{parriott2008,bregman2009}. The evolved stellar population of the galaxy is expected to contribute $\sim1~M_{\odot}$~yr$^{-1}$ per $10^{11}~M_{\odot}$ \citep{canning2013}, which means that at the current mass loss rate the amount of gas within the inner 10~kpc of MRK~1216 could be built up in less than 1 Gyr. The surface brightness and density profiles in MRK~1216 do not show the presence of abrupt jumps, indicating that the atmosphere of this isolated galaxy is continuous. We note that various parts of the hot gaseous atmosphere, with increasing radius, could in principle be labeled as inter-stellar medium (ISM), circum-galactic medium (CGM), and intergalactic medium (IGM). 

The presence of an X-ray atmosphere with a short nominal cooling time and the lack of young stars indicate the presence of a sustained heating source, which prevented star formation for 13 Gyrs, since the quick dissipative formation of the galaxy. The current Type Ia supernovae rate in the old stellar population of $2\times10^{11}~M_{\odot}$ is around 0.16 per 100 yr \citep{maoz2012}. This rate would be insufficient to balance the radiative cooling of the hot atmosphere of MRK~1216 even if the entire explosion energy of $10^{51}$ ergs per supernova would go into the heating of the X-ray emitting gas. The heating rate would fall short by at least an order of magnitude. Assuming that the supernova rate is decreasing with cosmic time as $(t/13\rm{Gyr})^{-1}$, already 1 Gyr after the formation of the stellar population of the galaxy, supernovae would not have been able to balance the radiative cooling of the hot atmosphere that today surrounds MRK~1216.

The central temperature peak, the relatively flat entropy profile \citep[index of $0.78\pm0.05$, which is flatter than the value of $\sim1.1$ expected from  gravitational collapse;][]{voit2005} and the presence of a radio source in the core of the galaxy indicate that, similarly to cooling core clusters and giant ellipticals, the heating source is radio-mechanical AGN feedback \citep[for review see][]{mcnamara2007,mcnamara2012}. The cooling time over free-fall time $t_{\rm cool} / t_{\rm ff}\sim15-30$ within the central $r\lesssim10$~kpc of MRK~1216 is also similar to the values measured in the centres of many clusters, groups and giant elliptical galaxies \citep{hogan2017,pulido2017}. The radially decreasing temperature profile is consistent with compressional heating in a steep central gravitational potential in the presence of sustained gentle heating \citep[see e.g.][]{gaspari2012b}. 
Similarly to the brightest cluster galaxies of luminous cooling core clusters \citep{larrondo2011,russell2013} and nearby giant ellipticals \citep{werner2012,werner2014}, the AGN is accreting at a very small X-ray luminosity $L_{\rm{X}}< 9.47\times10^{39}$~erg~s$^{-1}$, corresponding to an Eddington ratio of $L_{\rm{X}}/L_{\rm{Edd.}}\lesssim10^{-8}$. The radio luminosity is also similar to the giant ellipticals with ongoing radio-mechanical AGN feedback. The number of  counts is unfortunately too small to find AGN blown X-ray cavities in the X-ray images.

The black hole mass of $M_{\rm BH}=(4.9\pm1.7)\times10^9~M_{\odot}$ in MRK~1216 is a factor of 10 larger than the expectations from the black hole mass--bulge mass relation established at $z=0$ \citep{walsh2017,ferre2017}, but is within the intrinsic scatter measured by \citet{saglia2016} for the black hole mass--stellar velocity dispersion relation. PGC~032873 possibly also hosts an over-massive black hole, with a mass upper limit of $M_{\rm BH} < 10^{10}~M_{\odot}$ \citep{ferre2015}. The fraction of the current mass that these black holes already reached during the early stages of the fast dissipative growth $\sim13$ Gyr ago remains unknown. 
In light of the lack of significant mergers and star-formation in these cold gas free galaxies over the past 13~Gyr, the remaining black hole mass could only have been accreted from the hot halo. \citet{gaspari2017} show that the accretion rate onto the supermassive black hole is tightly linked to the X-ray properties of the hot halo, since it is the progenitor source of the feeding mechanism. 
In a physical process known as chaotic cold accretion (CCA; \citealt{gaspari2013}), multiphase gas condenses out of the hot halo \citep{gaspari2017b}, raining onto the central region, and via inelastic collisions the cold/warm clouds are rapidly funnelled toward the central black hole. 
High-resolution hydrodynamic simulations \citep[e.g.][]{gaspari2012} show that an isolated elliptical galaxy undergoing cycles of CCA and mechanical AGN feedback can grow a BH mass of several $10^9~M_{\odot}$ over 10 Gyr 
($M_{\rm BH} = E_{\rm inj}/(\varepsilon c^2)$, with a mechanical efficiency $\varepsilon = 10^{-4}$ and total injected mechanical energy of several $10^{59}$~erg). It is important to note that during CCA the galaxy experiences rapid flicker noise variability in time, thereby a currently low AGN X-ray luminosity does not necessarily imply low accretion rates throughout the cosmic time. Moreover, the nuclear radiative luminosity is expected to be low during the maintenance mode of feedback, since most of the power is released through the mechanical channel \citep[e.g.][]{russell2013} and the radiative efficiency strongly decreases with declining Eddington ratios. Finally, other observational studies have suggested that the total halo mass \citep[which in the case of Mrk~1216 is $M_{200}=9.6\pm3.7\times10^{12}~M_{\odot}$;][]{buote2017} correlates better with the black hole mass than the stellar populations of the host galaxies \citep{bogdan2012,bogdan2015,bogdan2018}. The study of the X-ray emitting gas in massive relic galaxies is therefore key to advance our knowledge of AGN feeding and feedback at the different evolutionary stages of giant ellipticals.

Given that over the past $\sim13$ Gyr both MRK~1216 and PGC~032873 appear to have evolved passively and in isolation, the difference in their current X-ray luminosity can perhaps be traced back to a difference in the ferocity of the AGN outbursts in these systems. The AGN activity in PGC 032873 could have been less gentle than the current activity in MRK~1216, undergoing a powerful outburst displacing most of the gas from its gravitational potential well.  A more detailed knowledge of the size of the scatter in the X-ray luminosities of isolated, massive, compact, relic galaxies will be obtained by the future optical and X-ray surveys, which are expected to discover more of these interesting systems in our cosmic backyard. 

\section{Conclusions}
\label{conclusions}
Here we report the first\footnote{About two weeks after submitting this paper and publishing its preprint on the arXiv, a preprint of an independent and complementary X-ray study appeared by \citet{buote2017}, focused mostly on studying the mass profile of MRK~1216.} detection of hot X-ray emitting atmospheres around the isolated, massive, compact, relic galaxies MRK~1216 and PGC~032873. For MRK~1216 within $r<10$~kpc, we find a 0.5--7~keV X-ray luminosity of $L_{\rm X}=(7.0\pm0.2)\times10^{41}$~erg~s$^{-1}$  and a gas mass of $M_{\rm gas}=(1.4\pm0.06)\times10^{9}~M_{\odot}$, similar to the local X-ray bright giant ellipticals many of which reside in the centres of groups.
\begin{itemize}
\item The galaxy shows a high central density of $n=0.30\pm0.02$~cm$^{-3}$, a central temperature peak of $kT=1.25\pm0.07$~keV, a short cooling time of  $t_{\rm cool}=52\pm5$~Myr, and a $\sim13$ Gyr old stellar population.  The presence of an X-ray atmosphere with a short nominal cooling time and the lack of young stars indicate the presence of a sustained heating source, which prevented star formation since the dissipative formation of the galaxy 13 Gyrs ago. The central temperature peak, the presence of radio emission in the core of the galaxy, and the low Eddington ratio indicate that the heating source is a gentle radio-mechanical AGN feedback.
\item The X-ray luminosity of the other isolated, massive, compact, relic galaxy PGC~032873, which has a similar stellar mass to MRK~1216 \citep{ferre2017}, is about an order of magnitude lower, $L_{\rm X}=(5.6\pm0.5)\times10^{40}$~erg~s$^{-1}$. Given that over the past $\sim13$ billion years both galaxies appear to have evolved passively and in isolation, the difference in their current X-ray luminosity could be traced back to a difference in the vigour of the AGN outbursts in these systems. 
\item MRK~1216 and possibly PGC 032873 are outliers above the black hole mass--bulge mass relation at $z=0$. The fraction of the current mass that these black holes already reached during the early stages of the fast dissipative growth $\sim13$ Gyr ago remains unknown. Theoretical predictions of chaotic cold accretion indicate that appreciable fractions of the black hole masses could have been accreted from the extended hot X-ray emitting halos. The study of the X-ray emitting gas in massive relic galaxies is thus key to advance our knowledge of AGN feeding and feedback at the different evolutionary stages of massive galaxies.
\end{itemize}

\section*{Acknowledgments}
We are grateful to the referee for the comments and suggestions (including the suggestion to add PGC~032873) which helped to improve the paper significantly.  
This work was supported by the Lend\"ulet LP2016-11 grant awarded by the Hungarian Academy of Sciences. M. G. thanks Fabrizio Brighenti for insightful discussions. M. G. is supported by NASA through Einstein Postdoctoral Fellowship Award Number PF5-160137 issued by the Chandra X-ray Observatory Center, which is operated by the SAO for and on behalf of NASA under contract NAS8-03060. Support for this work was also provided by Chandra grant GO7-18121X. 

\bibliographystyle{mnras}
\bibliography{clusters}

\end{document}